\documentclass[fleqn,usenatbib]{mnras}
\usepackage{newtxtext,newtxmath}
\usepackage[T1]{fontenc}
\usepackage{ae,aecompl}
\usepackage{graphicx}
\usepackage{amsmath}
\usepackage{subfig}
\usepackage{float}
\usepackage{threeparttable}
\usepackage{comment}
\restylefloat{figure}
\usepackage{url}
\usepackage{hyperref}
\usepackage{soul}
\sethlcolor{yellow} 
\usepackage{xcolor}

\def\lum{erg~s$^{-1}$}
\def\xa{XRT~000519}
\def\xb{XRT~110103}

\def\arcsec{\hbox{$^{\prime\prime}$}}
\def\approxlt{\ifmmode \rlap{$<$}{}_{{}_{{}_{\textstyle\sim}}} \else%
$\rlap{$<$}{}_{{}_{{}_{\textstyle\sim}}}$\fi}

\usepackage{hyperref}
\usepackage{academicons}
\usepackage{xcolor}
\usepackage{orcidlink}

\title[Candidate host for \xa]{Probing for the host galaxies of the fast X-ray transients XRT~000519 and XRT~110103}

\author[]{D. Eappachen\,\orcidlink{0000-0001-7841-0294}$^{1,2}$\thanks{E-mail: d.eappachen@sron.nl},
P.~G.~Jonker\,\orcidlink{0000-0001-5679-0695 }$^{2,1}$, 
M. Fraser\,\orcidlink{0000-0003-2191-1674}$^{3}$, 
M.A.P. Torres$^{4,5}$,\orcidlink{https://orcid.org/0000-0002-5297-2683}
V. S. Dhillon$^{6,4}$, 
\newauthor T.~Marsh$^{7}$, 
S. P. Littlefair$^{6 }$
J. Quirola-V\'asquez\,\orcidlink{0000-0001-8602-4641}$^{8,9,2}$ 
K. Maguire\,\orcidlink{0000-0002-9770-3508}$^{10}$, 
D. Mata Sánchez\,\orcidlink{0000-0003-0245-9424}$^{4,11,5}$,
\newauthor G. Cannizzaro$^{2,1}$, 
 Z. Kostrzewa-Rutkowska$^{12,1}$,
T. Wevers\,\orcidlink{0000-0002-4043-9400}$^{13,14}$, 
F. Onori\,\orcidlink{0000-0001-6286-1744}$^{15}$,  
\newauthor Anne Inkenhaag$^{2,1}$, 
S.J. Brennan$^{3}$
\\
\\
$^{1}$SRON, Netherlands Institute for Space Research, Niels Bohrweg 4, 2333 CA, Leiden, The Netherlands\\
$^{2}$Department of Astrophysics/IMAPP, Radboud University Nijmegen, P.O. Box 9010, 6500 GL, Nijmegen, The Netherlands\\
$^{3}$School of Physics, O’Brien Centre for Science North, University College Dublin, Belfield, Dublin 4, Ireland\\
$^{4}$Instituto de Astrof\'{i}sica de Canarias, E-38205 La Laguna, S/C de Tenerife, Spain\\
$^{5}$ Departamento de Astrof\'isica, Univ. de La Laguna, E-38206 La Laguna, Tenerife, Spain \\
$^{6}$Department of Physics \& Astronomy, University of Sheffield, Sheffield S3 7RH, UK\\
$^{7}$Department of Physics, Gibbet Hill Road, University of Warwick, Coventry, CV4 7AL, UK\\
$^{8}$Instituto de Astrof\'isica, Pontificia Universidad Cat\'olica de Chile, Casilla 306, Santiago 22, Chile\\
$^{9}$Millennium Institute of Astrophysics (MAS), Nuncio Monse$\tilde{n}$or S\'otero Sanz 100, Providencia, Santiago, Chile\\
$^{10}$School of Physics, Trinity College Dublin, the University of Dublin, College Green, Dublin 2, Ireland. \\
$^{11}$Jodrell Bank Centre for Astrophysics, School of Physics and Astronomy, The University of Manchester, M13 9PL, UK \\
$^{12}$Leiden Observatory, Leiden University, PO Box 9513, NL-2300 RA Leiden, The Netherlands\\
$^{13}$European Southern Observatory, Alonso de Cordova 3107, Vitacura, Santiago, Chile\\
$^{14}$Institute of Astronomy, Madingley Road, Cambridge CB3 0HA, United Kingdom \\
$^{15}$ INAF-Osservatorio Astronomico d'Abruzzo, via M.Maggini snc, I-64100 Teramo, Italia 
}
\date{Accepted XXX. Received YYY; in original form ZZZ}

\begin{document}
\label{firstpage}
\pagerange{\pageref{firstpage}--\pageref{lastpage}}
\maketitle

\begin{abstract}
 Over the past few years, $\sim$30 extragalactic fast X-ray transients (FXRTs) have been discovered, mainly in {\it Chandra} and {\it XMM-Newton} data. Their nature remains unclear,  with proposed origins including a double neutron star merger, a tidal disruption event involving an intermediate-mass black hole and a white dwarf, or a supernova shock breakout. A decisive differentiation between these three promising mechanisms for their origin requires an understanding of the FXRT energetics, environments, and/or host properties. We present optical observations obtained with the Very Large Telescope for the FXRTs \xa{} and \xb~and Gran Telescopio Canarias observations for \xa{} designed to search for host galaxies of these FXRTs. In the $g_s$, $r_s$ and $R$-band images, we detect an extended source on the North-West side of the $\sim$ $1^{\prime\prime}$ (68\% confidence) error circle of the X-ray position of \xa{} with a Kron magnitude of  $g_s=$26.29$\pm$0.09 (AB magnitude). We discuss the \xa{} association with the probable host candidate for various possible distances, and we conclude that if \xa\ is associated with the host candidate a supernova shock breakout scenario is likely excluded. No host galaxy is found near \xb{} down to a limiting magnitude of $R>25.8$.
\end{abstract}



\section{Introduction}
Fast X-ray transients  (FXRTs)  last a few tens to a few thousands of seconds. Over the past few years, FXRTs have been discovered in {\it Chandra}, {\it XMM Newton} and {\it eROSITA} data by \citet{2013jonker}; \citet{glennie}; \citet{Irwin}; \citet{Bauer}; \citet{Xue2019}; \citet{ATel2019}; \citet{Alp2020}; \citet{Novara2020}; \citet{2020ATel13416....1W}; \citet{2021ATel14599....1L}, although they were probably detected in data from earlier satellites as well (e.g., \citealt{2003ApJ...586.1238A, 2008Natur.453..469S}). 

The origin of fast X-ray flares may well be diverse: they have been proposed to arise from a binary neutron star merger (BNS), a tidal disruption event (TDE) involving an intermediate-mass black hole (IMBH)  and a white dwarf (WD), an off-axis or sub-luminous $\gamma$-ray burst (GRB) or a supernova shock breakout (SBO; see the references above and \citealt{2016ApJ...823..113P}; \citealt{2017hsn..book..967W}; \citealt{Xue2019};  \citealt{dadoanddar}). Potentially, the TDE of a planet-sized object by a neutron star in the halo of our Milky Way galaxy can explain the observed properties  (\citealt{2011Natur.480...69C}; \citealt{2016ApJ...823..113P}), similar to earlier models for $\gamma$-ray bursts. 


"Shock breakout" is the earliest electromagnetic radiation from a supernova explosion. The SBO emission takes place when the radiation--mediated shock reaches the edge of the star (\citealt{2010ApJ...724.1396O}; \citealt{2010ApJ...725..904N}; \citealt{2017hsn..book..967W}). Brief X-ray flares can arise in a supernova SBO. A famous example of such a shock breakout signal is that discovered serendipitously by  {\it Swift}  for SN~2008D (\citealt{2008Natur.453..469S}). Prompt discovery of supernovae and the properties of the shock breakout X-ray emission allow for the radius of the progenitor star to be inferred (\citealt{2017hsn..book..967W}), which is crucial to set constraints on supernova simulations. 
More observational input especially during the first phase of the explosion are needed to provide new constraints to the physics governing these simulations. 
Among the FXRTs reported by \citet{Alp2020}, nine of them are associated with potential host galaxies and the FXRTs XT~070618, XT~060207 and XT~040610 are host-less candidates.  \citet{Alp2020} explain all these FXRTs as supernova shock breakout emission. 

Tidal disruption of stars was predicted over three decades ago by e.g. \citet{1975Natur.254..295H} and \citet{1988Natur.333..523R}. TDEs release a large amount of energy as bright flares at X-ray, ultra-violet  (UV)  and optical wavelengths  (\citealt{2016Sci...351...62V}; \citealt{2020SSRv..216...85S}). Because the tidal radius and Schwarzschild radius scale differently with black hole  (BH)  mass, solar-mass stars cross the event horizon before being disrupted for supermassive black hole  (SMBH)  masses exceeding $\sim$ $10^{8}$ M$_\odot$. There would be no observable signals except possibly for gravitational waves in such cases   (\citealt{1989A&A...209..103L}; \citealt{2014ApJ...795..135E}). Note that this mass limit can be higher in the situation where the SMBH is rapidly spinning and the star follows an orbit prograde with this spin (\citealt{1992MNRAS.259..209B}; \citealt{2016NatAs...1E...2L}), or when the to-be-disrupted star is less compact than a 1~M$_\odot$ main-sequence star such as a more massive main-sequence star or a giant star (where the outer envelope can be disrupted; \citealt{2013ApJ...767...25G}). 

In the case of a TDE one also expects a shock breakout signal (e.g., \citealt{2009ApJ...705..844G,2019MNRAS.487.4083Y}). Detection of this tidal disruption X--ray breakout signal would be important as it would allow the measurement of the time-delay between the disruption of the star and the onset of the release of power for instance through a self-interaction shock or through accretion of gas on to the black hole detectable for instance through optical and X-ray emission. This could settle the debate on the nature of the optical emission (shock powered or accretion powered: \citealt{1988Natur.333..523R}; \citealt{2015ApJ...806..164P}). The time delay between onset of optical and X-ray emission also contains valuable information on the TDE properties (\citealt{2021arXiv210703666H, 2021ApJ...914...69C}).

The properties of several FXRTs are similar to those predicted for the TDE of a white dwarf (WD)  by an  IMBH (\citealt{2009ApJ...695..404R}; \citealt{2020SSRv..216...39M}). We define IMBHs as those black holes that have a mass between 100 M$_\odot$ and $10^6$ M$_\odot$   (cf.~\citealt{Greene2019arXiv191109678G}). WDs can be tidally disrupted by IMBHs with masses below $\sim$ $10^{5}$ M$_\odot$   (again this mass limit depends on the black hole spin; \citealt{2020SSRv..216...39M}). Globular clusters and dwarf galaxies are considered promising locations to host IMBHs  (\citealt{1999ApJ...519...89C}; \citealt{2001ApJ...554.1035F}; \citealt{2004ApJ...616..221G}; \citealt{2007ApJ...670...92G}), although the WD TDE rate is thought to be about 1/100 of the main-sequence TDE rate around IMBHs (10$^{-6}$ per globular cluster per year e.g., \citealt{2016ApJ...819...70M, 2020SSRv..216...39M}). 

A number of attempts have been made to identify the host galaxies for FXRTs in optical data. The fast X-ray transient CDF-S XT2 was found to be associated with a galaxy at a redshift $z$~=~0.74. The clear plateau in its X-ray light curve, similar to those seen in short GRBs, argues for a BNS origin. The FXRT lies in the outskirts of the star forming host galaxy with an offset  of $\approx$3.3 $\pm$ 1.9 kpc from galaxy centre (\citealt{Xue2019}). This is consistent with the host properties of short $\gamma$-ray bursts (\citealt{2014ARA&A..52...43B}). \citet{Xue2019} and gives a possible explanation for the FXRT  as being powered by a millisecond magnetar, formed in the aftermath of a binary neutron star merger (\citealt{2008MNRAS.385.1455M}).  The light curve of the FXRT XRT~210423 (\citealt{2021ATel14599....1L}) is similar to that of CDF-S XT2, hence it also has been proposed as the result of a binary neutron star merger (\citealt{2021ApJ...915L..11A}).

CDF-S XT1 is found to be associated with a faint host galaxy of uncertain redshift (\citealt{Bauer}). Serendipitous VLT imaging observations were obtained only 80 min after the X-ray detection of FXRT CDF-S XT1 by \citet{Bauer}. No optical counterpart was found down to a limiting magnitude of $\sim$25.7 in the $R$-filter. If the lack of a bright optical counterpart close in time to the detection of the X-ray transient is common, it is essential to study their host galaxy properties inorder to determine the distance and thus energetics of these FXRTs, .

In this paper we report on late-time optical observations designed to find or constrain the host galaxy properties of \xa{} (\citealt{2013jonker}) and \xb{} (\citealt{glennie}). \xa{} lies in  the direction  of  M86. If  we assume it is at {the distance of M86 of 16.2 Mpc}, this X-ray transient has a peak luminosity of $\sim$ 6$\ \times \ 10^{42}$ \lum. The X-ray light curve shows that the main flare is double-peaked, with precursor events taking place approximately 4000~s and 8000~s before the main flare (\citealt{2013jonker}). The precursor event timescale is in agreement with the expected orbital timescale of a white dwarf in an eccentric orbit around an IMBH   (\citealt{2016ApJ...819...70M}). The observed tail in the \textit{Chandra} X-ray light curve of \xa{} would then be associated with the accretion of part of the material falling back   (at super-Eddington rates) towards the IMBH   (\citealt{2009MNRAS.400.2070S}; \citealt{2011MNRAS.410..359L}). 

The X-ray transient \xb~if associated with the ACO~3581 cluster {at a distance of 94.9 Mpc }as proposed by \citet{glennie}, has a peak luminosity of $\sim$2 $\times$ $10^{44}$  \lum. For 6560~s of observations with \textit{Chandra} prior to the main flare there was no detection of the source above the background.  

\xa~ is located at right ascension (R.A.) and  declination (Dec) $12^{\rm h} 25^{\rm m} 31^{\rm s}.64$,  $+13^{\rm \circ} 03^{\rm \prime} 58^{\rm \prime\prime}.8$ while the best known coordinates for \xb~are $14^{\rm h} 08^{\rm m} 28^{\rm s}.89$,  $-27^{\rm \circ} 03^{\rm \prime} 29^{\rm \prime\prime}.4$ (J2000).
The 1-$\sigma$ uncertainty in the source position is $1^{\prime\prime}$ and $1.1\arcsec{}$ for \xa{} and \xb{}, respectively (from \citealt{2013jonker,glennie}).

Throughout the paper, we have considered the base $\Lambda$-CDM cosmology, with Hubble constant  H$_0$= 67.4$\pm$0.5 km/s/Mpc, matter density parameter $\Omega_m$=0.315$\pm$0.007 (\citealt{2018arXiv180706209P}). In \S~2 we describe the observations and our analysis, in \S~3 we list the results, which we discuss in \S~4, and we conclude in \S~5.


 \section{OBSERVATIONS AND ANALYSIS}
A journal of the photometric observations of the FXRTs \xa{} and \xb{} is given in Table \ref{tab:Table_Xtrans}.

 \begin{table*}
\small
\begin{center}
\caption{A journal of the photometric observations of the X-ray transients \xa{} and \xb{} used in this paper. }
\label{tab:Table_Xtrans}
\begin{tabular}{llllllll}
\hline
Target &  Telescope/Instrument  & Date & Observations & {Filters} & Exposure[s] & Airmass & Seeing \\
		\hline
		\xa{} &  VLT/FORS2 &  2018 Feb 12 & 3 & R & 1050 & 1.3 & $\sim 0.90^{\prime\prime}$\\
		\xb{} & VLT/FORS2 & 2018 Mar 23, 26 & 3 $+$ 3 & R & 1050 & 1.0 & $\sim 0.96^{\prime\prime}$, $0.90^{\prime\prime}$\\
		\xa{} &  GTC/HiPERCAM &  2021 May 10 & 17 & $u_s$, $g_s$, $r_s$, $i_s$, $z_s$& 180 & 1.1 & $\sim 0.8^{\prime\prime}$\\
\hline
\end{tabular}
\end{center}
\end{table*}

 \subsection{VLT observations of \xa\,and \xb{}}
We obtained optical images of the fields around FXRT \xa{} and \xb{} using the European Southern Observatory (ESO) 8.2~m Very Large Telescope (VLT) employing the FOcal Reducer/low dispersion Spectrograph  (FORS2; \citealt{1998Msngr..94....1A}).
It is equipped with two 2k$\times$4k MIT CCDs. 
We only used CCD1 for our analysis as, by design, the fields around FXRT \xa{} and \xb{} were covered by that CCD. The images have been taken in the Johnson-Cousins $R$-band filter which has an effective wavelength $\lambda_{{\rm eff}}  $ of 640~nm with a bandwidth $  (\Delta\lambda) $ of 158~nm   (\citealt{2005ARA&A..43..293B}). We took 3 $\times$ 1050~s images for \xa{} and 6 $\times$ 1050~s for \xb. The initial plan was to observe both the sources for 3 $\times$ 1050~s, however, due to deteriorating seeing over the first observations of \xb{}, they were repeated under better seeing conditions. The observations were done in service mode.

We performed bias and flat field correction using the ESO reflex   (\citealt{2013A&A...559A..96F})  data reduction pipeline. The $\textsc{L.A.Cosmic}$ software was employed to remove cosmic rays from the images  (\citealt{Lacos2001PASP..113.1420V}). We examined the masks for each of the images and made sure that none of the pixels around the fields of our interest has been masked erroneously.

After cosmic ray removal, the individual frames were aligned and  average combined for both \xa{} and \xb. The  {\textsc{IRAF IMCOMBINE}} task was used to stack the images. For \xa{} we only used the second and  third images for stacking as the seeing (FWHM) for the first image was $1.14^{\prime\prime}$, while the other two images were substantially better with values of 0.88 and 0.90$^{\prime\prime}$. Similarly, for \xb{}, we only used five images for the deep stacked image, as the seeing of the third image was worse ($0.99^{\prime\prime}$), compared to an average of $0.91^{\prime\prime}$ for the other five images.

In order to put the stacked images of both sources on the International Celestial Reference System   we considered all the Pan-STARRS sources with number of detections (ndetections)$>$5, within $3^{\prime}$ of the telescope pointing centre. We manually inspected all of the sources on the FORS2 image and excluded any source that was either saturated or extended. We measured the pixel coordinates in the FORS2 image of all the remaining sources using the centroid algorithm in {\textsc IRAF PHOT}. Finally, we used the {\textsc IRAF CCMAP} task to determine the new World Coordinate System (WCS) solution. Sixty-nine and fifty sources were used to fit this astrometric transformation for \xa{} and \xb{}, respectively. The {\textsc IRAF CCSETWCS} task
was used to apply the astrometric calibration to the images.

Next, we extracted the R.A. and Dec, magnitude, and magnitude error from all the objects detected in the stacked image using the Source Extractor software ($\textsc{SExtractor}$; \citealt{1996A&AS..117..393B}). For the photometric calibration of the image we used Pan-STARRS catalogue data  (\citealt{2016arXiv161205560C}). The transformation equation of \citet{Lupton:2005} was used to convert the Pan-STARRS magnitudes given in the Sloan $r_s$- and $i_s$ bands to a Johnson-Cousins $R$-band magnitude.

Finally, we compared the instrumental magnitude of each of the stars extracted using $\textsc{SExtractor}$ with its magnitude in the Pan-STARRS catalogue excluding saturated stars. The median of the difference between the instrumental and the Pan-STARRS magnitudes was calculated. This gives us the zero-point value. 
We used the calculated zeropoints for \xa\,and \xb\,separately to convert the instrumental magnitudes to apparent magnitudes in both cases.
We used Python version 3.7, with Astropy packages for the further analysis  (\citealt{astropy:2013}; \citealt{astropy:2018}). For any galaxy near \xa{} and \xb\,the Kron magnitudes were calculated.

\subsection{GTC observations of \xa}

We obtained simultaneous $u_s$, $g_s$, $r_s$, $i_s$, and $z_s$-band images with the HiPERCAM instrument mounted on the 10.4~m Gran Telescopio Canarias (GTC) at the  Roque de los Muchachos Observatory (La Palma, Spain). HiPERCAM makes use of four dichroic beamsplitters to split the collimated light into five filters.  Seventeen frames with an exposure time of 180~s each with only 7.8 ms dead time between each exposure were obtained on May 10, 2021 starting at 23:26:40 (UTC) in each of the filters. HiPERCAM provides a plate scale of 0.081\arcsec~per pixel and a field of view of 3.1$^\prime$(diagonal) on GTC (\citealt{hipercam2021MNRAS.507..350D}). The seeing in the images was around 0.8\arcsec. 

The data reduction steps including bias subtraction, flatfield correction and, for the $z_s$-band, fringe correction, which were performed using a dedicated data reduction pipeline\footnote{In particular, we used the {\sc joinup} script. See https://deneb.astro.warwick.ac.uk/phsaap/hipercam/docs/html/ Fringe frames obtained in May 19, 2018 were applied. These are available at https://deneb.astro.warwick.ac.uk/phsaap/hipercam/docs/html/files.html\#fringe-maps-and-peak-trough-pairs}. We checked the individual images for significant seeing or sky transparency variations but these were not present. We used the $\textsc{L.A.Cosmic}$ software to remove cosmic rays from the individual images  (\citealt{Lacos2001PASP..113.1420V}). Then, we average--combined the 17 images into one deep image.

\begin{figure*} 
    \centering
	\includegraphics[height=3.7cm, width=\textwidth]{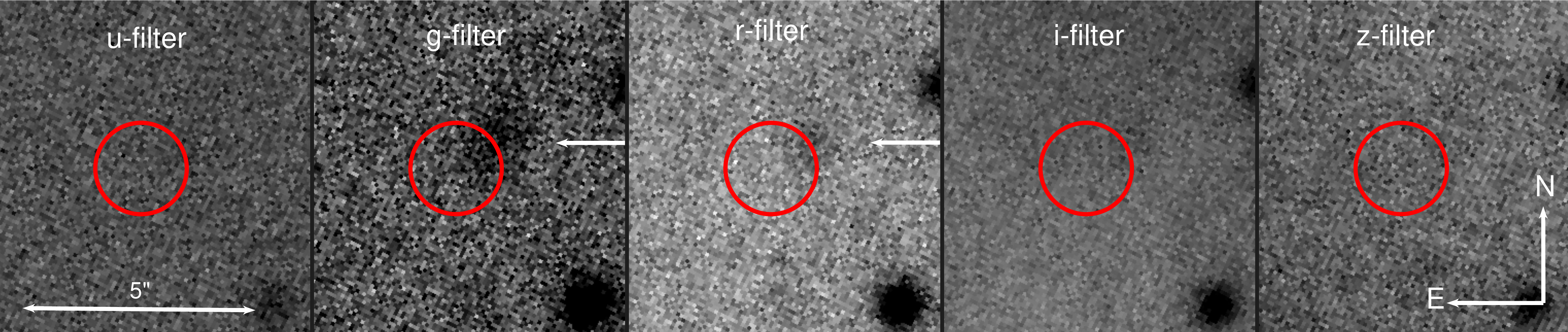}
    \caption{The GTC/HiPERCAM images of the field of \xa{}. From left to right: the $u_s$, $g_s$, $r_s$, $i_s$, and $z_s$-band images, respectively. The 68\% confidence circular error region with a radius of  $1^{\prime\prime}$ is shown by a red circle. The candidate host galaxy situated to the North-West of the position of \xa\,is marked with a white arrow in the $g_s$ and $r_s$ bands.}
    \label{fig:hipercam}
\end{figure*}

For each of the combined $u_s$, $g_s$, $r_s$, $i_s$, and $z_s$-band images, we refined the default astrometric solution that was based on the telescope pointing using the known astrometric position of 11 stars in the Pan-STARRS catalogue. We used the centroid algorithm in {\textsc IRAF PHOT} and then {\textsc IRAF CCMAP} to determine the WCS solution. For each filter we obtained independent astrometric solutions and we applied the astrometric corrections using {\textsc IRAF CCSETWCS} task to the corresponding filters.   The HiPERCAM images of the $\sim$ 5$^{\prime\prime}$ $\times$ 5$^{\prime\prime}$ region of \xa{} are given in the Figure \ref{fig:hipercam}. For the zero point value, we compared the instrumental magnitude of each of the stars to its magnitude in the Pan-STAARS catalogue for the $g_s$, $r_s$, $i_s$, and $z_s$-filters and  to the SDSS catalogue for the $u_s$-filter. 

\subsection{WHT observations \xa}
 We obtained spectra of {six nearby galaxies} to try to determine their redshift using Auxiliary-port CAMera (ACAM) at the 4.2-m William Herschel Telescope (WHT). Some information on the observed galaxies is given in Table~\ref{tab:Table_gal}. We used the V400 grism, the CG395A (transmitting 3950–9400 \AA{}) order blocking filter, the 1\arcsec~slit, and the AUXCAM CCD for our observations. Arc lamp spectra were obtained after each science spectrum exposure. The spectra were reduced using \textsc{PyRAF}  (\citealt{2012ascl.soft07011S}) and {\sc molly}  (\citealt{2019ascl.soft07012M}). 

\begin{table*}
\small
\caption{Selected information on galaxies near \xa. The number in the first column corresponds to the number of the galaxy in Figure~\ref{fig_gal}. {The first four galaxies may belong to a group of galaxies at $z=0.18$.} }
\label{tab:Table_gal}
\begin{tabular}{ccccccccc}
\hline
\multicolumn{1}{|p{0.5cm}|}{\centering Galaxy}
&\multicolumn{1}{|p{2cm}|}{\centering Name}
&\multicolumn{1}{|p{2cm}|}{\centering R.A.~and Dec \\ (J2000)}
&\multicolumn{1}{|p{2cm}|}{\centering Apparent Kron $R$ magnitude } 
&\multicolumn{1}{|p{1cm}|}{\centering  Kron radius \\ ($^{\prime\prime}$) } 
&\multicolumn{1}{|p{2cm}|}{\centering redshift}
&\multicolumn{1}{|p{2cm}|}{\centering Projected size (kpc) }
&\multicolumn{1}{|p{2cm}|}{\centering  Absolute Magnitude} \\
 \hline
1 & SDSS J122535.17+130411.4 & $12^{\rm h}25^{\rm m}35^{\rm s}$  $+13^{\rm \circ}04^{\rm \prime}12^{\prime\prime}$ & 20.04 $\pm$ 0.02 & 3.51 & $0.1866 \pm 0.0001$ & 11.4 & -19.83 $\pm$ 0.03  \\
 2 & SDSS J122532.20+130501.9 & $12^{\rm h}25^{\rm m}32^{\rm s}$  $+13^{\rm \circ}05^{\rm \prime}03^{\prime\prime}$ & 19.15 $\pm$ 0.02 & 3.89 & $0.1866  \pm 0.0001$  & 12.6 & -20.72 $\pm$ 0.03\\
 3 & SDSS J122546.74+130440.4 & $12^{\rm h}25^{\rm m}46^{\rm s}$  $+13^{\rm \circ}04^{\rm \prime}40^{\prime\prime}$ & 19.37 $\pm$ 0.02& 3.5 & $0.1866  \pm 0.0001$ & 11.3 &-20.50 $\pm$ 0.03 \\
 4 &SDSS J122540.24+130445.7 & $12^{\rm h}25^{\rm m}40^{\rm s}$  $+13^{\rm \circ}04^{\rm \prime}45^{\prime\prime}$ & 18.82 $\pm$ 0.02 & 3.5 & $0.1866  \pm 0.0001$ & 11.3 &-21.05 $\pm$ 0.03 \\
\hline

 5 & SDSS J122539.24+130714.1 & $12^{\rm h}25^{\rm m}39^{\rm s}$ $+13^{\rm \circ}07^{\rm \prime}14^{\prime\prime}$ & 18.80 $\pm$ 0.02 & 3.5 &$0.29 \pm 0.05$ & 15.7 & -22.14 $\pm$ 0.04 \\
  6 & SDSS J122536.71+130219.0 & $12^{\rm h}25^{\rm m}36^{\rm s}$ $+13^{\rm \circ}02^{\rm \prime}19^{\prime\prime}$ & 17.52 $\pm$ 0.01$^{\ast}$ & 7.5 & $0.1082 \pm 0.0012$ & 15.3 & -21.06 $\pm$ 0.01 \\
 \hline
\end{tabular}

\begin{flushright}
$^{\ast}$ magnitude from SDSS
\end{flushright}
\end{table*}

\subsection{Completeness and limiting magnitude}
For all VLT and GTC images we determine the completeness and limiting magnitude through the addition of artificial stars close to the location of the X-ray source. We define the completeness as the magnitude at which 95 per cent of injected artificial stars are being recovered at $>$5~$\sigma$ and within 0.2 magnitude of the input magnitude, while we define the magnitude at which 33 per cent of the artificially created stars are recovered as the limiting magnitude. 

In detail the procedure is as follows: using {\textsc{IRAF}}'s {\textsc {PSF}} task we first create a model for the point spread function (PSF) using isolated stars in the image. We next add an artificially created star of variable magnitude with a light distribution consistent with the PSF to the relevant image. We used a \textsc{PyRAF} script to add the artificial star to the image using the {\textsc ADDSTAR} task. The artificial star is added to the 68\% confidence $1^{\prime\prime}$ transient uncertainty region associated with the X-ray detection. Within this region we varied the position where we put the artificial star on the image slightly to sample variations in the background count rate. 

We tried to detect this artificially created star using the standard source detection algorithm used before ({\textsc DAOFIND}; with a 5$\sigma$ detection threshold) and we also ran {\textsc PHOT}, for the sources which are being detected. The artificial star is counted as recovered if the source magnitude is determined to within 0.2 magnitude of the input value. 

We repeat the above procedure for artificial stars with different magnitudes in 0.1 steps in magnitude. We first determined the relevant magnitude range using 100 iterations. After the important range in magnitude is found we iterated the artificial star procedure for 10$^3$ times for each 0.1 magnitude bin.

\section{RESULTS}
\subsection{\xa: Candidate host identification and photometry}

\begin{figure} 
    \centering
	\includegraphics[height=7.5cm, width=8.3cm]{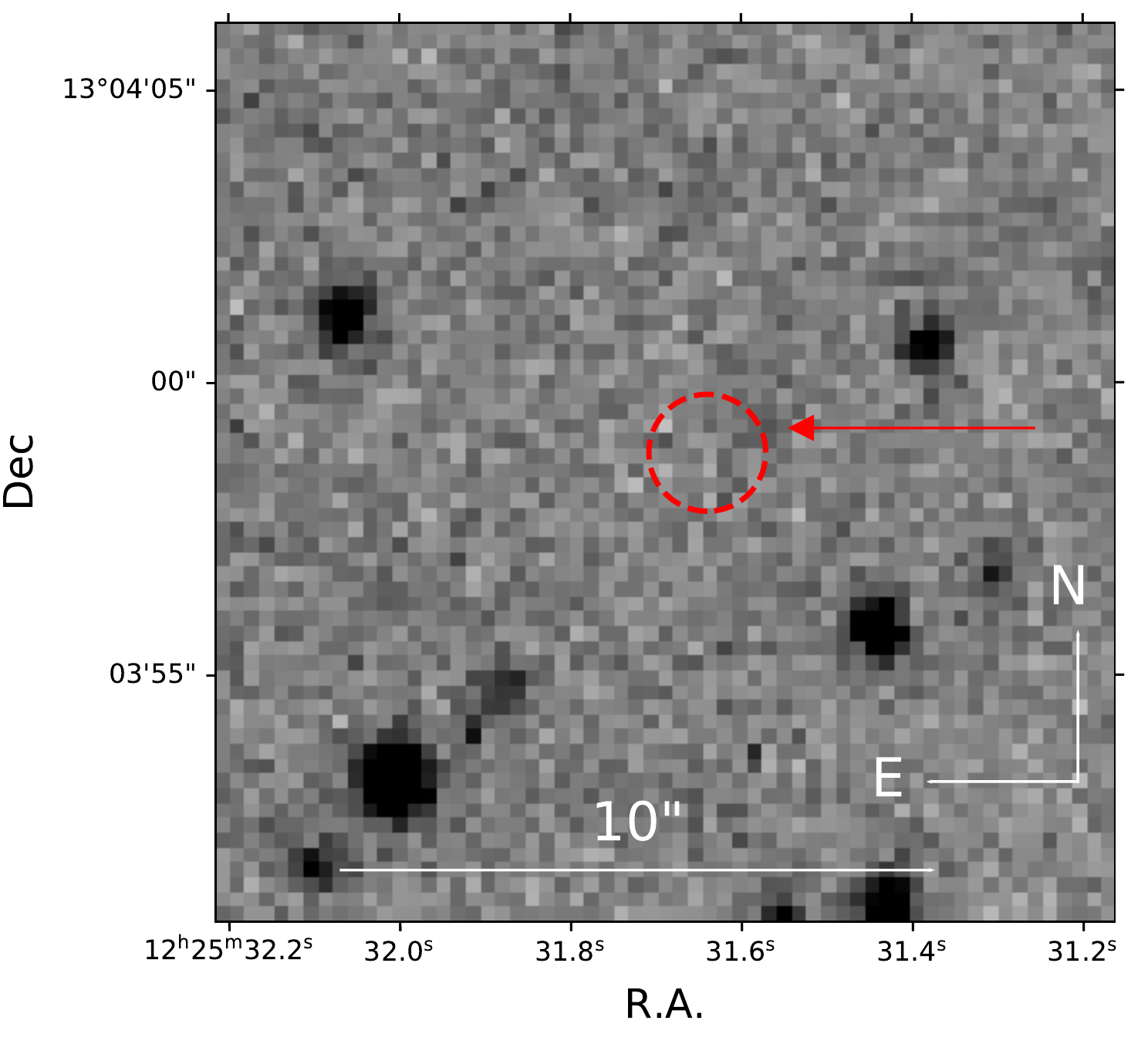}
    \caption{The VLT/FORS2 image of the  $\approx$10\arcsec$\times$10\arcsec\,field around the best-known position of \xa{} is shown. The 68\% confidence circular error region with a radius of  $1^{\prime\prime}$ is marked by a red circle.  The marginally detected candidate host to the North-West of the position of \xa\,is marked with a red arrow. 
    }
    \label{fig:519}
\end{figure}

{We clearly detect a source (>5$\sigma$) (called cNW hereafter) in the GTC/HiPERCAM $g_s$ and $r_s$--band images on the North-Western side of the $\sim$ $1^{\prime\prime}$ error circle of the \xa{} position at R.A.~and Dec~$12^{\rm h}25^{\rm m}31.58^{\rm s}$,  $+13^{\rm \circ}03^{\rm \prime}59.32^{\prime\prime}$(see Figure}~\ref{fig:hipercam}). {Our marginal detection (3$\sigma$) in VLT/FORS2 $R$-band observations corroborates this} (see Figure ~\ref{fig:519}).


From the HiPERCAM images, we obtained a Kron magnitude ({\textsc MAG\textunderscore AUTO}) for the probable host galaxy cNW  by using the Source Extractor software ($\textsc{SExtractor}$; \citealt{1996A&AS..117..393B})  { on the $g_s$-filter image $m_{g_s}$=26.29$\pm$0.09.}   

\begin{figure} 
    \centering
	\includegraphics[height=4.0cm, width=9.0cm]{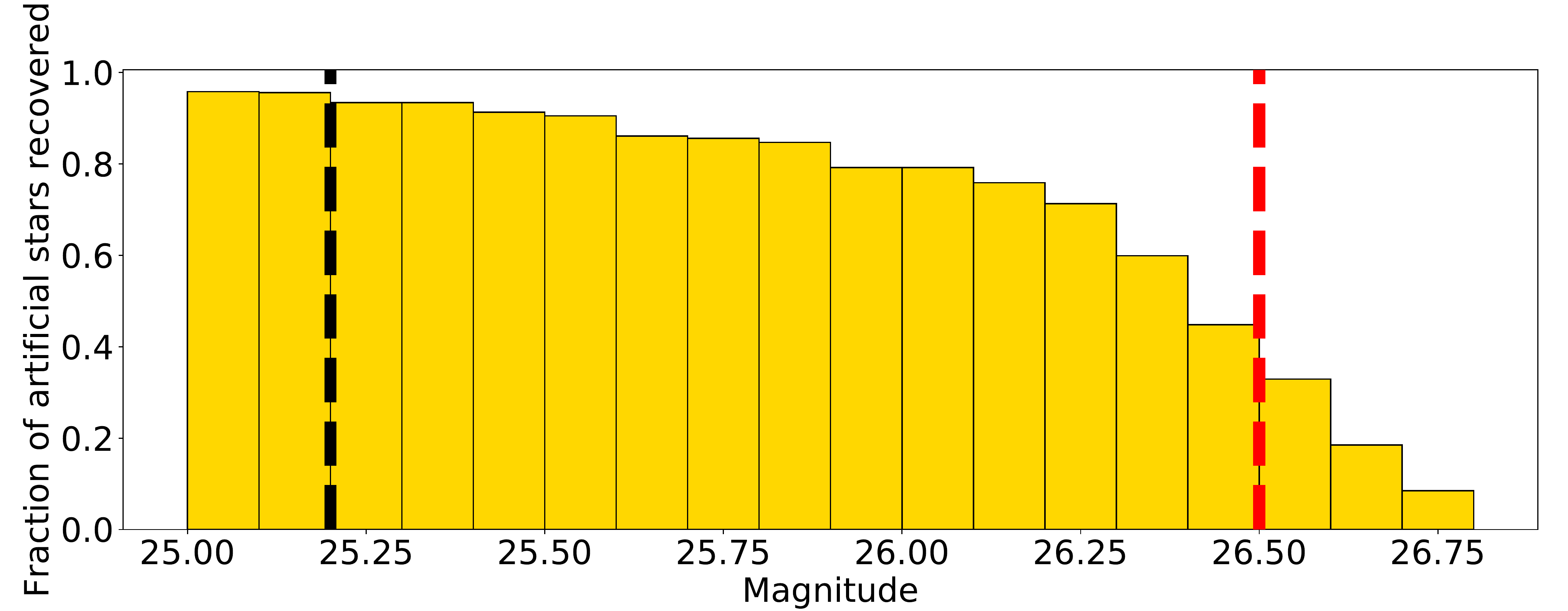}
    \caption{{The completeness limit (dashed black line at 25.2) and the limiting magnitude (red dashed line at 26.5) for the HiPERCAM $g_s$-band image of \xa{} are shown.}}
    \label{fig:df_g}
\end{figure}

We also determined the completeness and limiting magnitude for $u_s$, $g_s$, $r_s$, $i_s$, and $z_s$-band  HiPERCAM images. {They are shown in Figure~\ref{fig:df_g} for the HiPERCAM $g_s$-band image of \xa{}  (see \S~2.4 for details)}. The completeness limits are   $u_s$=24.6, { $g_s$=25.2, $r_s$=25.0,} $i_s$=24.5, and $z_s$=24.0, whereas the limiting magnitudes are $u_s$=26.0,  {$g_s$=26.5, $r_s$=25.8}, $i_s$=25.6, and $z_s$=25.3. 

We also employ {\sc GALFIT}  (\citealt{2010AJ....139.2097P}) to fit the light distribution of cNW in the $g_s$-- and $r_s$--band images to try and constrain the galaxy morphology. We use the PSF model that we created from bright and isolated stars with the {\textsc IRAF} {\textsc SEEPSF} task in the respective filters as input for {\sc GALFIT}. A de Vaucouleur profile ({\it S\'ersic} with index 4) gives the best fit (reduced ${\chi}^2$ = 1.019; degrees of freedom [d.o.f.]=3018) for the $g_s$-band image.  We could not determine the galaxy morphology in the $r_s$-filter. This is probably caused by the low number of galaxy photons detected in the $r_s$ filter image. 

From the $g_s$-band {\sc GALFIT} fit we obtain an effective half-light radius  ($R_{e}$) of $\sim$8.68 pixels. With a pixel scale of 0.081$^{\prime\prime}$ per pixel for HiPERCAM on GTC, cNW has an angular size for $R_{e}$ of 0.703$^{\prime \prime}\pm0.002^{\prime\prime}$. Given the seeing of 0.8$^{\prime\prime}$, this implies that the half-light diameter of the galaxy is {resolved}. {We also obtained the $g_s$-- $r_s$ colour of cNW from the HiPERCAM image using aperture photometry in  $\textsc{SExtractor}$  obtaining $g_s$-- $r_s=-0.11 \pm$ 0.19 ($g_s$ = $26.16 \pm 0.11$ and $r_s$ = $26.27 \pm 0.16$, using a circular aperture of $1^{\prime\prime}$  radius). Note that the colour is not corrected for any inter-galactic/stellar extinction.}

We determine the probability for cNW to be found inside the 68$\%$ confidence region of \xa~by chance. First, we determine the number density of sources brighter than or as bright as cNW from the FORS image. It is 0.028 stars/sq.~arcsec in a region of $30^{\prime\prime}$ $\times$ $30^{\prime\prime}$ centered on R.A.~$12^{\rm h} 25^{\rm m} 31^{\rm s}.2$ and Dec.~$+13^{\rm \circ} 03^{\rm \prime} 57^{\rm \prime\prime}.2$. This region is chosen such that it falls away from the stream of stars (see \citealt{2013jonker}). {Considering that the $1^{\prime\prime}$ error region has an area of 3.14 sq.~arcsec, and assuming Poisson statistics, 
we find the probability for one or more sources to be found randomly in the localization error region to be 0.083.}


\begin{figure} 
    \centering
	\includegraphics[height=9.8cm, width=9cm]{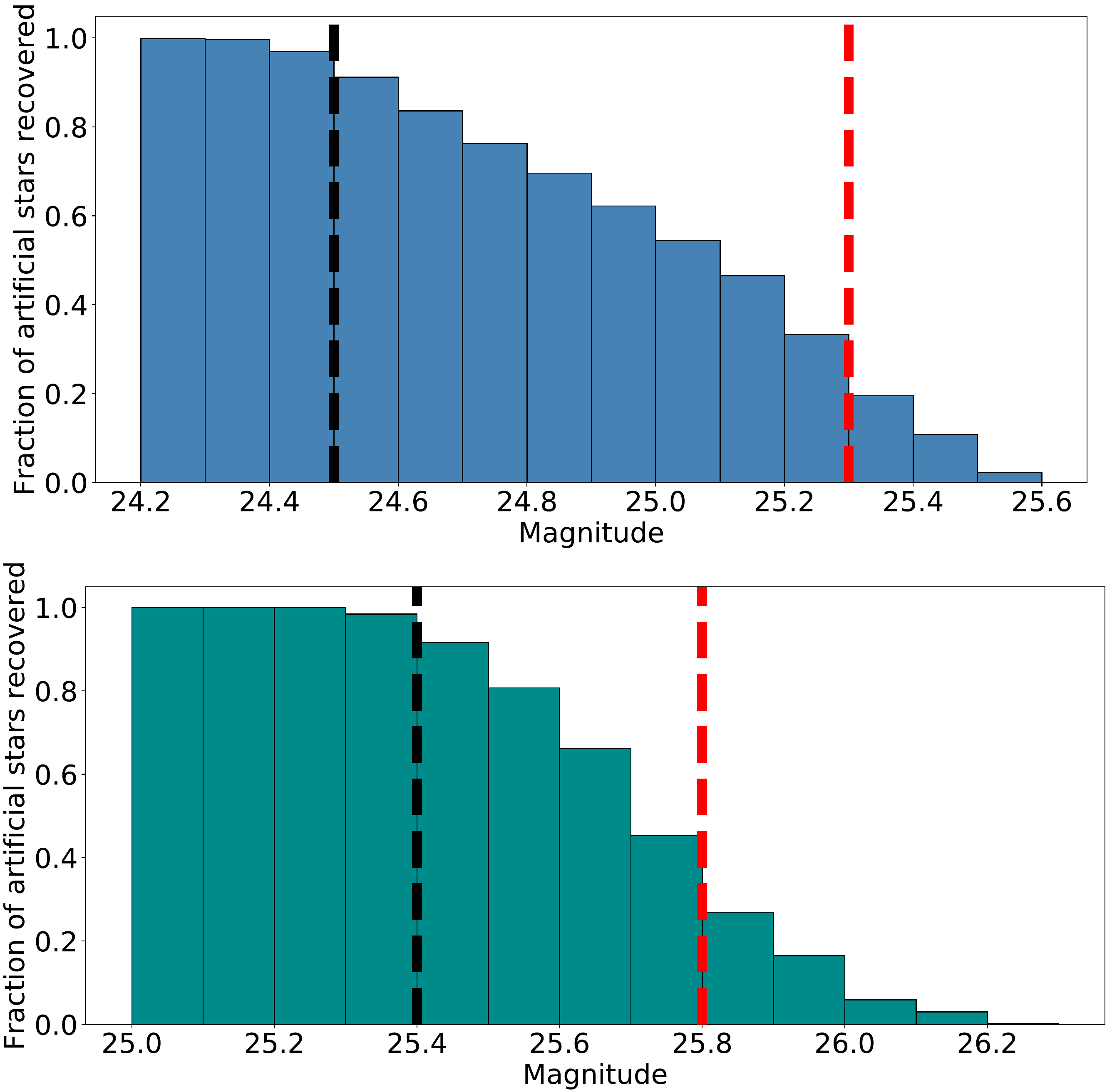}
    \caption{ The fraction of artificial stars that is recovered with a source magnitude lying within 0.2 magnitude of the input value (see text for details), for the VLT/FORS2 $R$-band images of \xa{} (top) and \xb{} (bottom). The dashed black lines indicate the completeness limit and the red dashed lines denote the limiting magnitude (see the section \S~2.4 for our definitions of these parameters). Note the different x-axis values for the two plots.}
    \label{fig:df}
\end{figure}

We used {\textsc{IRAF}} to obtain PSF photometry of cNW in the FORS $R$-band. The source cNW has a PSF magnitude of $R\sim$~26.0$\pm$0.3, whereas we derive a completeness limit of $R=$24.5 and limiting magnitude of $R=$25.3. {The completeness and limiting magnitude for the FORS2 $R$-band is derived through the procedure mentioned in \S~2.4. For \xa{} and \xb{} they are shown in Figure}~\ref{fig:df}. That the PSF magnitude is fainter than the completeness and limiting magnitude can be explained in part by the use of a 5~$\sigma$ detection threshold in our artificial star experiment, whereas the source cNW is only detected at a 3~$\sigma$ level in the FORS2 $R$-band image. Furthermore, as mentioned in \S~2.4 the limiting magnitude is defined as the magnitude at which 33 percent of the stars are recovered. This implies that the faintest source detected could be fainter than the defined limiting magnitude. {The magnitude for the object reported in} \citet{2013jonker} {is not that of the candidate host galaxy but instead that of a nearby high pixel.}

\subsection{\xa: Spectroscopy of Field Galaxies }
\begin{figure}
    \centering
	\includegraphics[height=4.2cm, width=8.2cm]{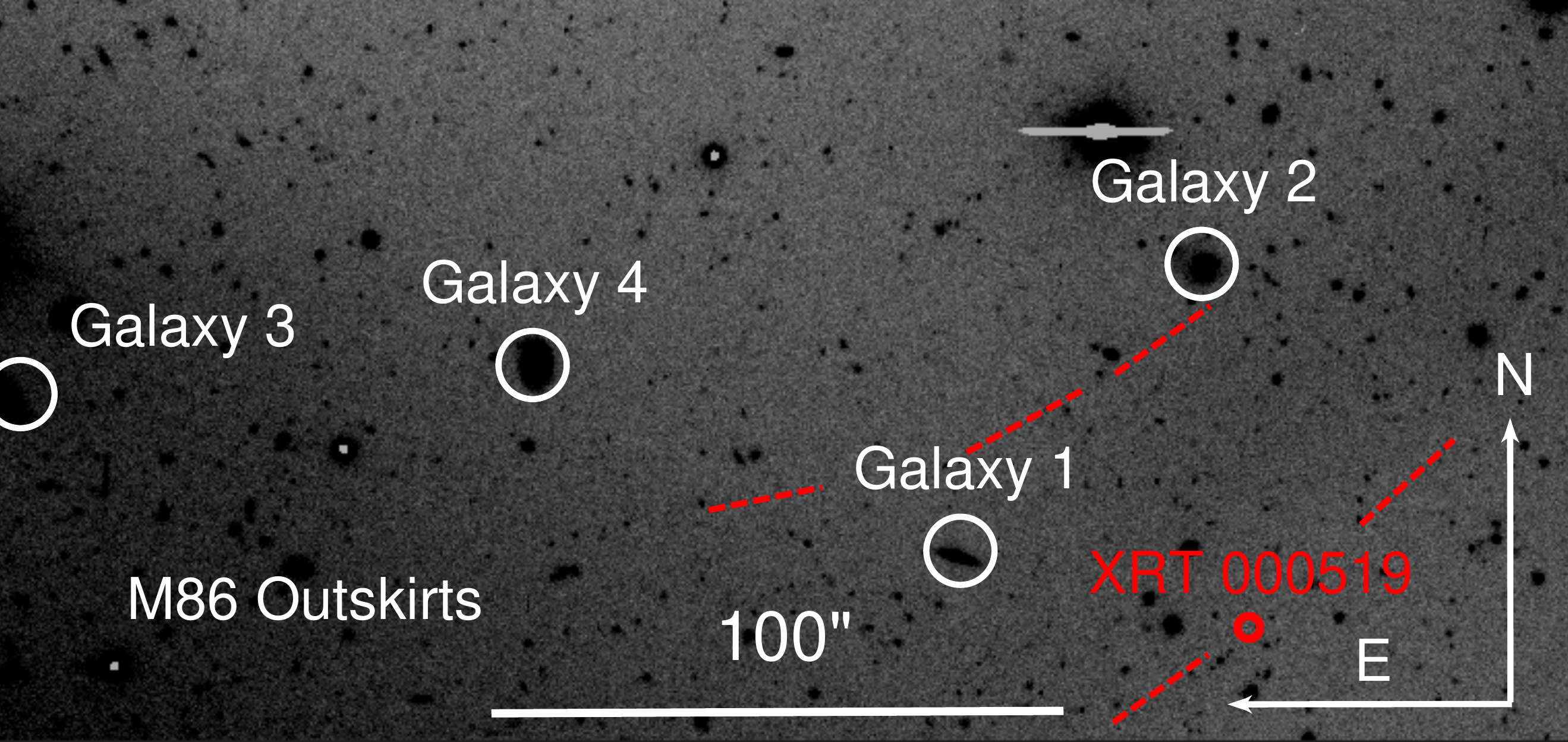}
    \caption{ VLT/FORS2 $R$-filter image of the field around the transient \xa{} (which is indicated with the red circle). The red dashed lines show the approximate location of the stream of stars, stripped from the galaxy SDSS~J122541.29+130251.2, which in this Figure is hidden in the brightness of the stars of the outskirts of M86. We also show the galaxies near \xa{} for which we took a spectrum in order to check the possibility of a galaxy group beyond M86.}
    \label{fig_gal}
\end{figure}

\begin{figure}
 \flushleft
	\includegraphics[height=4.5 cm,width=9 cm]{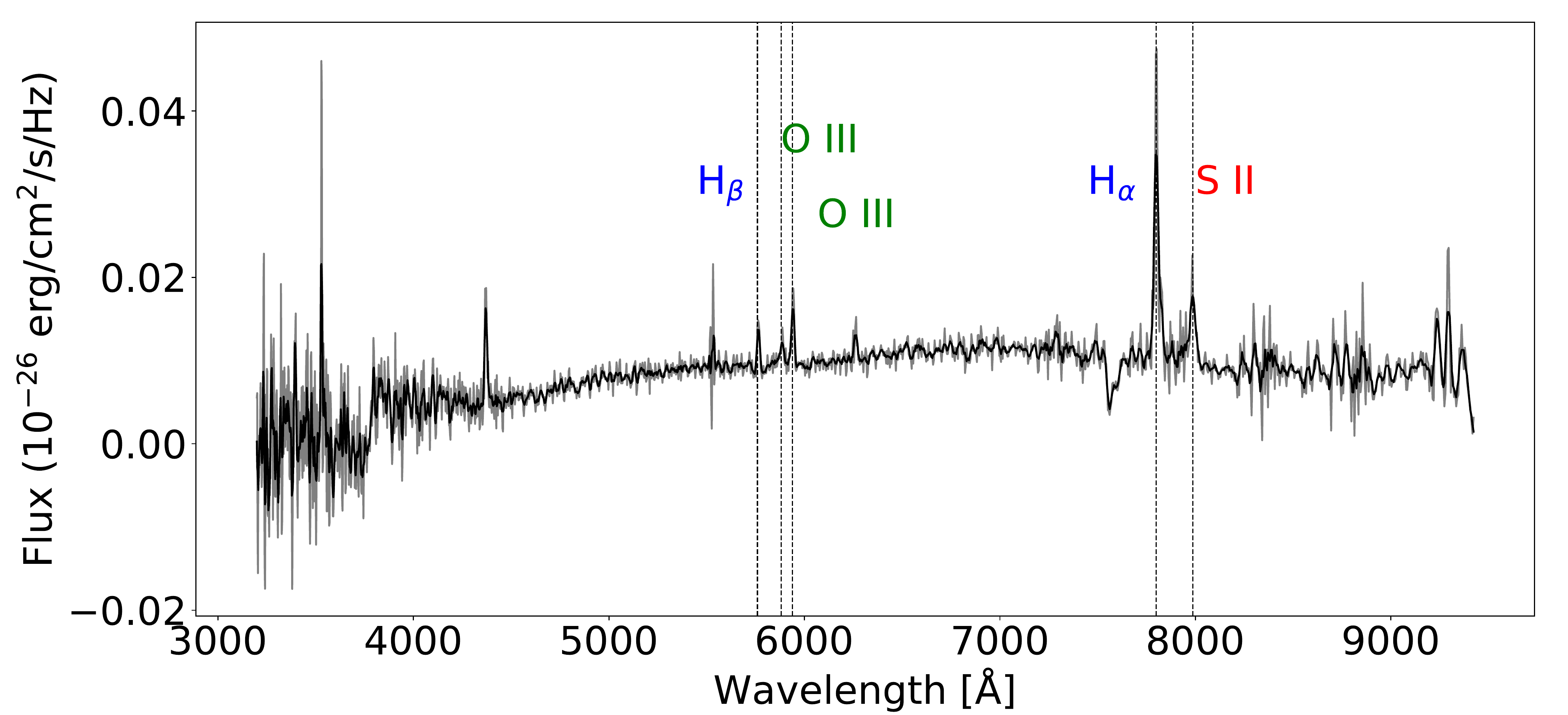}
    \caption{The flux calibrated spectrum of the galaxy SDSS~J122540.24+130445.7 (Galaxy 4 in Fig.~\ref{fig_gal}). Unbinned data is shown in gray. The boxcar smoothed spectrum (Box1DKernel with a width of 5 pixels) is shown in black. . The dotted lines indicate the emission lines in the spectrum that we used for the redshift determination ($z$ = 0.1866 $\pm 0.0001$).  }
    \label{spectrum}
\end{figure}

We obtained spectra of six galaxies in the field of \xa{} including four galaxies that form part of a possible group (Table \ref{tab:Table_gal}). Figure~\ref{fig_gal} {shows those four galaxies. For each we obtained a spectrum.} The redshift for the source labelled "Galaxy 4" was determined using  the emission lines H$\alpha$ $\lambda$6564, H$\beta$ $\lambda$4862, [O III] $\lambda$4960, 5008 and [Si II] $\lambda$6732. {The flux calibrated spectrum of "Galaxy 4"  is shown  Figure~\ref{spectrum}. We fitted multiple Gaussians to the emission lines using the {\sc lmfit} \footnote{https://lmfit.github.io/lmfit-py/} package and obtained the best-fit central wavelengths and their associated errors.  The redshift is $z = 0.1866  \pm 0.0001$, giving a luminosity distance of 940.2 $\pm$ 0.4 Mpc calculated using } \citet{2006PASP..118.1711W}. {We determined the redshift of "Galaxy 5" to be $0.29 \pm 0.05$ by cross-correlating the ACAM spectrum with the template spectra obtained by the Sloan Digital Sky Survey 5} (\citealt{2007ApJS..172..634A}). {Emission lines H$\alpha$ $\lambda$6564, [O III] $\lambda$4960, 5008 and [Si II] $\lambda$6732 were  present in the spectra of "Galaxy 6" and we determined the redshift to be $0.1082 \pm 0.0012$.  For the three galaxies labeled 1--3 in  Table~\ref{tab:Table_gal}, we cross-correlated the WHT spectra with those of different galaxy templates obtained by the Sloan Digital Sky Survey 5 in an attempt to determine their redshifts. However, no significant cross-correlation signal was found. In addition, no emission lines were detected in the spectra that could help us determine their redshift. }

\subsection{\xb{}}
\begin{figure} 
\centering
	\includegraphics[height=7.5 cm,width=8.6 cm]{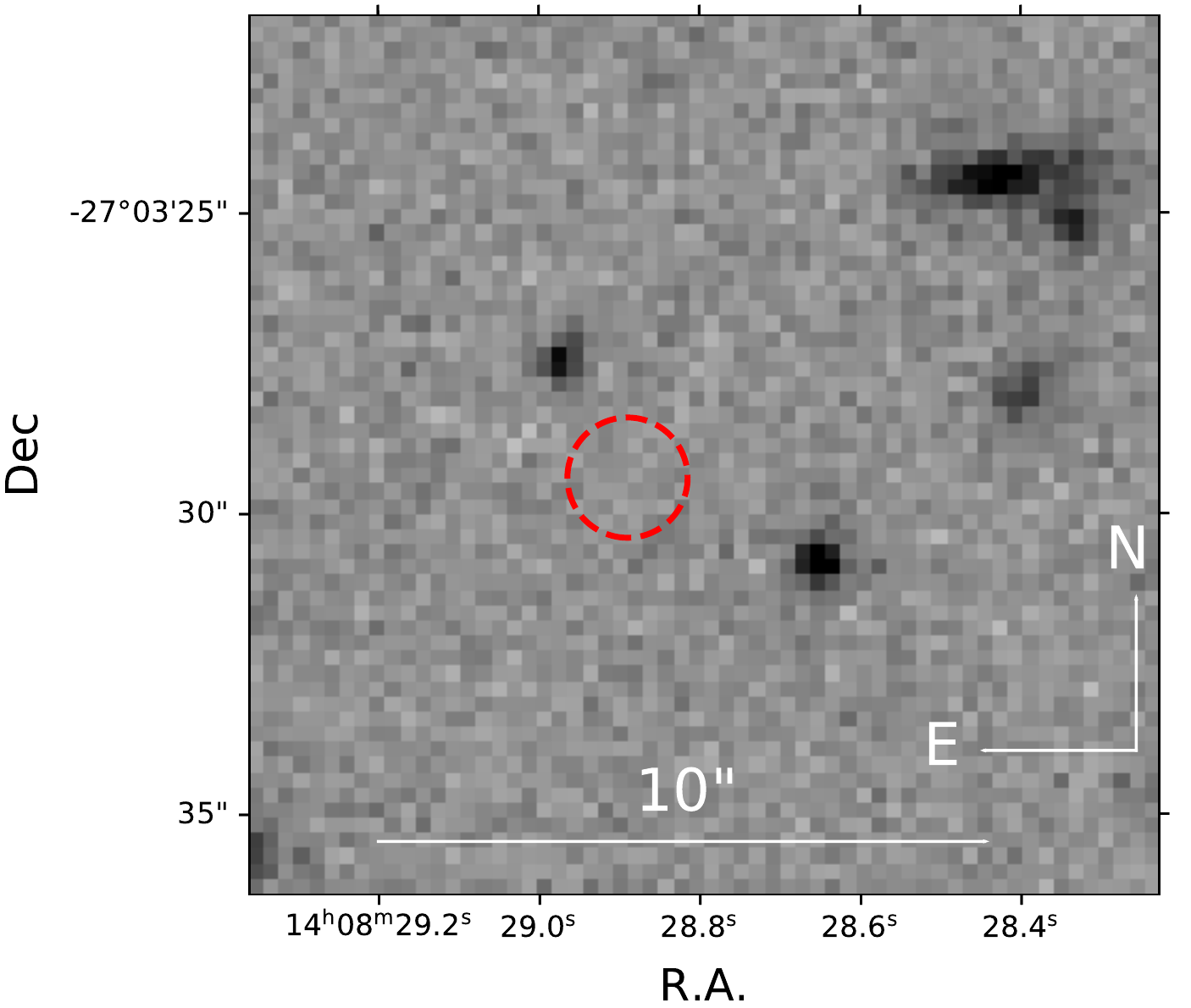}
    \caption{The VLT/FORS2 image of the $\approx$10\arcsec$\times$10\arcsec\,field around the best-known position of \xb{} is shown. The 68\% confidence circular error region with a radius of  $1^{\prime\prime}$ is marked by a red circle. }
    \label{fig:103}
\end{figure}

For \xb~ we find no candidate host in the positional error region (see Figure \ref{fig:103}) down to a completeness and limiting magnitude of $R=$25.4 and $R=$25.8, respectively (see Figure \ref{fig:df}). {The nearby sources in the image are consistent  with point sources as the PSF subtraction gave us clean residuals at the resolution of VLT/FORS.}

\section{Discussion}
We obtained deep optical images to try to detect the host galaxies of the FXRTs \xa~and \xb. The detection of a host galaxy or globular cluster would enable a redshift/distance measurement, which would allow the observed flux and fluence to be converted to luminosity and energy, respectively. These parameters would strongly constrain the nature of these FXRTs.
For example,  the peak luminosities of different progenitor models are drastically different, with $L_{\rm{X,peak}}\approxlt 10^{46}$ erg s$^{-1}$ for BNS mergers (\citealt{2014ARA&A..52...43B}), $L_{\rm{X,peak}} \approxlt 10^{48}$ erg s$^{-1}$ WD-IMBH TDEs (\citealt{2020SSRv..216...39M}), and $L_{\rm{X,peak}} \approxlt 10^{45}$ erg s$^{-1}$  for supernova SBOs (\citealt{2008Natur.453..469S}; \citealt{2017hsn..book..967W}). Furthermore, the small- and large-scale environment properties of the FXRT such as the host galaxy offset, and the host properties such as the stellar age and star-formation rate also provide important clues on the nature of these events.  

\subsection{\xa{}}

We have detected an extended source in the North-West region of the circular 95 per cent confidence error region of \xa. This source, which we call cNW, is visible in Figures~\ref{fig:hipercam} and ~\ref{fig:519}. The extended nature of the source is most readily detectable in the HiPERCAM $g_s$-band image. A {\sc GALFIT} fit gives a half light radius of $\approx$0.7$^{\prime\prime}$\footnote{In order to estimate the uncertainty on the half light radius we varied the input parameters for our {\sc GALFIT} fit, however, the resulting median absolute deviation on the half light radius was 0.002$^{\prime\prime}$, a value we deem too small to be a realistic estimate for the uncertainty. }.

\subsubsection{Assuming cNW and \xa~are associated}

\begin{figure}
    \centering
	\includegraphics[height=8.3cm, width=8.8cm]{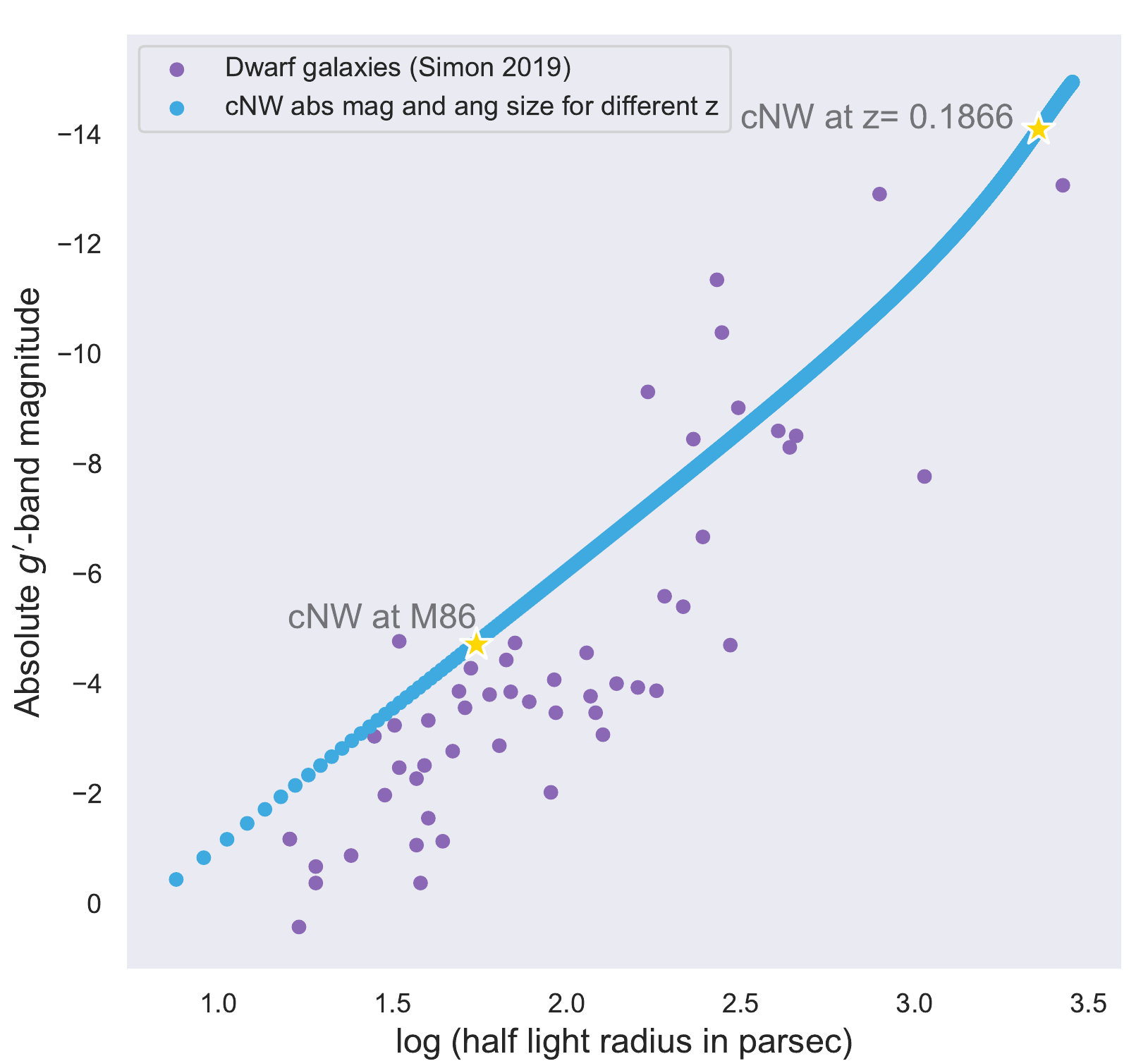}
    \caption{ The absolute $g_s$-band magnitude and half-light radius of dwarf galaxies is shown by purple circles (the data comes from \citealt{2019ARA&A..57..375S}). The absolute (K-corrected) $g_s$-band magnitude  and half light radius of the candidate host galaxy (cNW) calculated for different distances/redshifts is denoted by blue circles. Here we determined the K-correction using the K-correction calculator (\citealt{2010MNRAS.405.1409C}, \citealt{2012MNRAS.419.1727C}). The location of cNW assuming it were either at the distance of M86 or at $z$=0.1866 is marked with a yellow star symbol. For both distances the source falls close to the size-luminosity relation observed for dwarf galaxies, although it would either have to be brighter given its size or smaller given its absolute magnitude than typical if at the distance of M86.}
    \label{fig:dwarf}
\end{figure}

In order to investigate the nature of \xa\ we will first consider two possible distances to cNW. 
The fact that the projected position of cNW falls close to a stream of stars seen to protrude from M86 leads us to consider the possibility that it lies at the distance of M86. If so, it has an absolute magnitude of {$M_{g_s}=-4.8$. }At that distance the angular half light radius of $\approx$0.7$^{\prime\prime}$ converts to a projected half light radius of $\sim$55 pc. We compared this inferred size and absolute magnitude with the size and absolute magnitude distributions observed for dwarf galaxies in the $g^\prime$-band \footnote{For the purpose of the calculation in this section we assume that $g_s$ = $g^\prime$ and $r_s$ = $r^\prime$ } (see Figure \ref{fig:dwarf}). Here, we used the absolute magnitudes and half light radii of dwarf galaxies given in \citet{2019ARA&A..57..375S}.  To convert their $V$-band magnitudes to the $g^\prime$-band magnitudes we used the  transformation equation by \citet{Lupton:2005}. For the $g^\prime$ -- $r^\prime$ colour necessary for the transformation we took the median color of  50000 galaxies from the SDSS using the {\textsc AstroML sdss\textunderscore galaxy\textunderscore colours} function where {\textsc specClass = 'GALAXY'}. We calculated the absolute magnitude and half-light radius of cNW assuming different redshifts in Figure~\ref{fig:dwarf}, K-corrected using the
K-calculator\footnote{See http://kcor.sai.msu.ru/} (\citealt{2010MNRAS.405.1409C}, \citealt{2012MNRAS.419.1727C}). {We used the $g^\prime$ -- $r^\prime$ derived from the HiPERCAM images for the K-correction for cNW. } 

{For the inferred absolute magnitude, the inferred half light radius is small compared with that observed for dwarf galaxies, or alternatively, the source is bright for its size. The HiPERCAM images show that cNW is blue with $g^\prime$ -- $r^\prime=-0.11\pm0.19$. Given the colour of cNW, considering the absolute magnitude and angular size of blue compact dwarf galaxies (\citealt{2014A&A...562A..49M}, \citealt{2015llg..book..323K}), we deem it unlikely that cNW is a dwarf galaxy at the distance of M86. But we cannot definitively rule out the possibility of cNW being an ultra-faint dwarf galaxy at the distance of M86.}
A dwarf galaxy could harbour an IMBH (\citealt{2020SSRv..216...39M}, \citealt{2020ARA&A..58..257G}) and the X-ray flare could be due to an IMBH-WD TDE.  

If \xa~is at the distance of M86, the implied peak X-ray luminosity of $\sim$ 6$\ \times \ 10^{42}$ \lum , does in itself not rule out an SBO origin, given that it is consistent with the peak luminosity seen and expected in the case of supernova SBOs (\citealt{2008Natur.453..469S}; \citealt{2017hsn..book..967W}). In this scenario the supernova went off in a dwarf galaxy. However, it is quite unlikely that \xa{} is due to a supernova SBO at M86 as it is unlikely that the associated optical supernova was missed (see section 3.1 of \citealt{2013jonker}). 

{We next check if cNW, and by association \xa{}, is at the distance of the putative group of galaxies.}  From Figure~\ref{fig_gal} it can be seen that there are several galaxies (in projection) near the location of \xa. The redshift for galaxy~4 is $z = 0.1866  \pm 0.0001$, giving a luminosity distance of 940.2 $\pm$ 0.4 Mpc. Its associated absolute magnitude and angular/physical size are $-$21.05 and $\sim$ $3.5^{\prime\prime}$/11.3 kpc, respectively.
If the galaxies 1--3 are at the same redshift, the projected distance between the galaxies with the largest angular separation on the sky is $\sim$ 0.68 Mpc, and with {four galaxies}, it is reasonable to consider this a compact galaxy group (see e.g., \citealt{2021Univ....7..139L}).
If we assume that all {four galaxies} belong to a galaxy group beyond M86, we can calculate their absolute magnitudes (see Table~\ref{tab:Table_gal}). The spread in these absolute magnitudes is small ({mean $\sim-20.52$ and standard deviation $\sim$0.45}).  The detected angular size, again assuming they are at the same distance, also implies that their physical size are similar (Table~\ref{tab:Table_gal}).  We next assume that cNW lies at the same distance as this putative group of galaxies. This implies an absolute magnitude of {M$_{g_s}$ = $-$14.1 (K-corrected)} and a projected half light radius of 2.2 kpc for cNW (see Figure \ref{fig:dwarf}). This absolute magnitude and projected size make the source consistent with the (extrapolated) size-luminosity relation for dwarf galaxies (\citealt{2019ARA&A..57..375S}).  
{Several blue dwarf galaxies also have a similar size and absolute magnitude to that of cNW if cNW were to be at the distance of the probable galaxy group} (\citealt{2014A&A...562A..49M}, \citealt{2015llg..book..323K}). If \xa~is associated with such a background galaxy group at a distance of $\sim$940 Mpc the peak luminosity of the \xa{}  would be $\sim$2 $\times $10$^{46}$ erg s$^{-1}$. In that case, we could discard a supernova SBO origin as the progenitor (\citealt{2008Natur.453..469S}; \citealt{2017hsn..book..967W}), and if the source originates in a WD TDE or binary neutron star merger it implies that the peak luminosity is super-Eddington (perhaps as the result of beaming). { It is interesting to note that hydrogen-poor superluminous supernovae also seem to favour blue-star forming galaxies }(for a review see \citealt{2019ARA&A..57..305G}).

\begin{figure} 
    \centering
	\includegraphics[height=8.2cm, width=8.6cm]{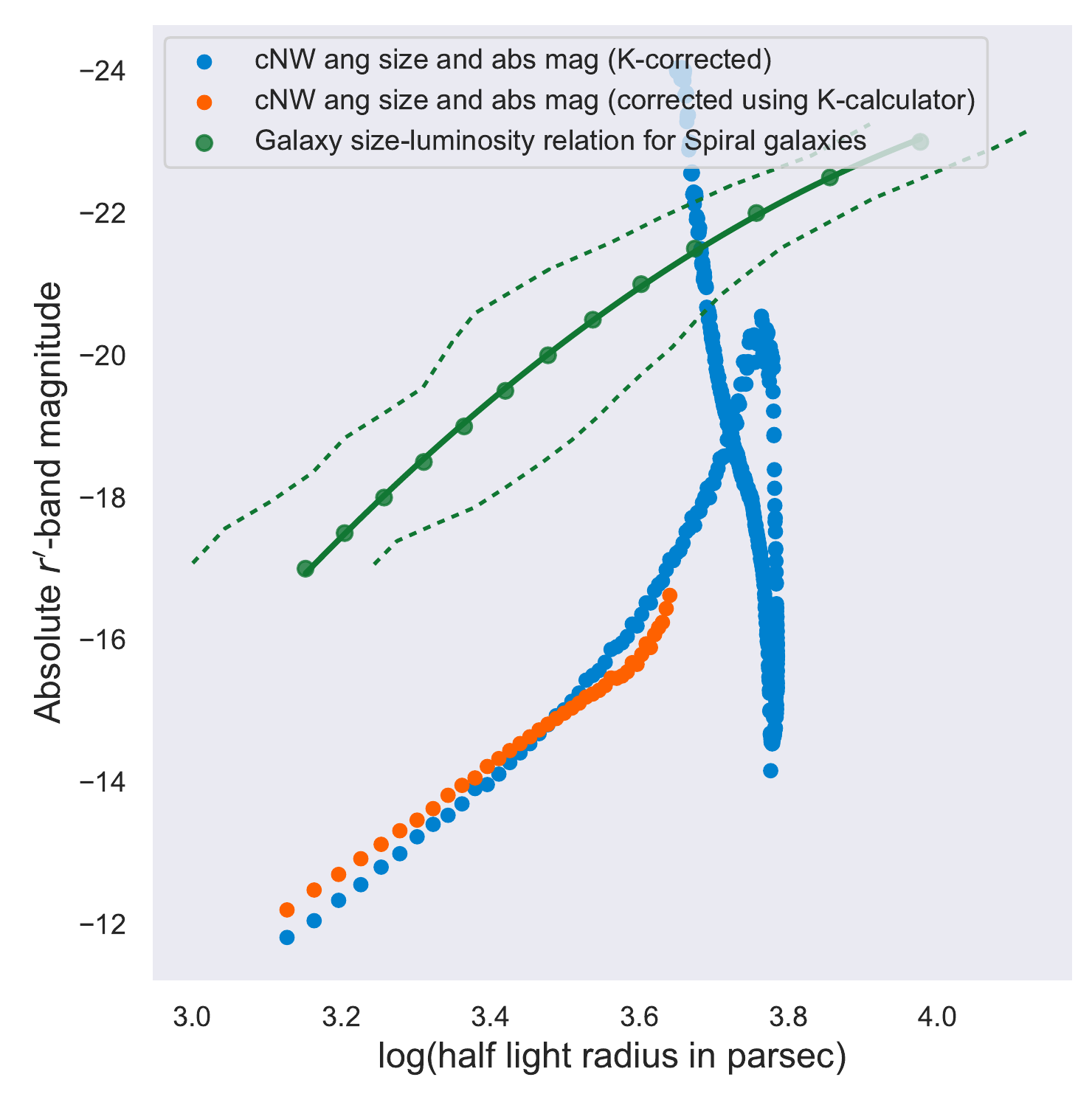}
    \caption{ For different redshifts, the blue dots show the absolute magnitude  in the $r_s$-filter and the half-light radius of cNW in parsec. We used \citet{2002astro.ph.10394H} to calculate the K-correction using the SED of the spiral galaxy. The orange points are determined using the K-correction calculator (\citealt{2010MNRAS.405.1409C}, \citealt{2012MNRAS.419.1727C}) which is valid for a $z<$0.5. The spiral galaxy size-luminosity relation is given in  green (\citealt{2019RAA....19....6Z}), along with the 1-$\sigma$ confidence region indicated by the region between the dotted lines. The two relations cross for a redshift in the range {of 4.1--4.5. }}
    \label{fig:spiral}
\end{figure}

\begin{figure}
    \centering
	\includegraphics[height=8cm, width=8.5cm]{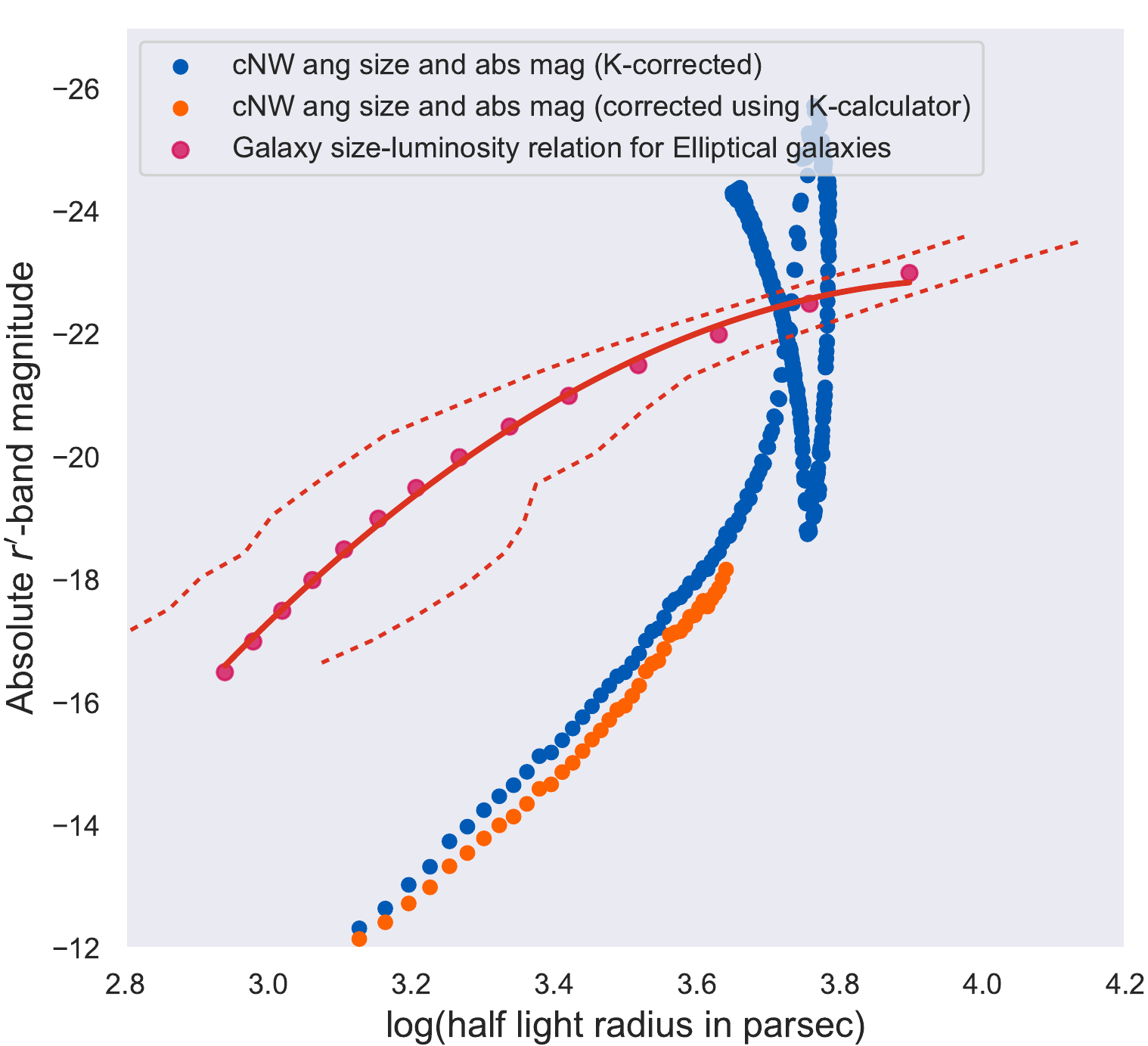}
    \caption{ The blue points show the absolute $r_s$-band magnitude and the half-light radius for cNW for varying redshifts where the K-correction is calculated using the prescription  given in \citet{2002astro.ph.10394H}. We used the SED of elliptical galaxy  for K-correction. The orange circles give the values for cNW using the K-correction calculator (\citealt{2010MNRAS.405.1409C}, \citealt{2012MNRAS.419.1727C}) which is valid for a $z<$0.5. The galaxy size-luminosity relation for elliptical galaxies is given in red (\citealt{2019RAA....19....6Z}). The 1-$\sigma$ confidence region on this relation is given by the region enclosed by the red dotted lines. The two relations cross for a redshift in the range of {0.75--0.80, 1.80--1.85, or 3.4--3.8}}
    \label{fig:ellip}
\end{figure}

{ Next, we investigate if we can obtain distance constraints on \xa~ interpreting cNW as a spiral or an elliptical galaxy. To this end, we compared the size and absolute (Kron) magnitude of cNW when placed at varying distances }to that of spiral and elliptical galaxies. For Figures \ref{fig:spiral} and \ref{fig:ellip},  we calculated the K-correction using the prescription presented in \citet{2002astro.ph.10394H} using the spectral energy distribution (SED) of a spiral or an elliptical galaxy to compute the K-correction. We used SED models given in the {\textsc HYPERZ} package\footnote{See http://www.bo.astro.it/$^\sim$micol/Hyperz/old\_public\_v1/hyperz\_manual1/ node6.html}(\citealt{2000A&A...363..476B}), {\small CWW\textunderscore Scd \textunderscore ext.sed} and {\small CWW\textunderscore E\textunderscore ext.sed} for spiral- and elliptical galaxies, respectively (see Figures \ref{fig:spiral} and \ref{fig:ellip}). {Note that for the {\sc HYPERZ} templates the original data extends only from $1400$ to $10000$\AA{}; the data in the templates is extrapolated at ultraviolet and near-infrared wavelengths using the spectral evolution models of \citet{1993ApJ...405..538B}. 
 In addition, we used the SED templates {\small elliptical\textunderscore template} and {\small s0\textunderscore template} from the {\textsc Kinney-Calzetti spectral atlas} (\citealt{1994ApJ...429..582C}, \citealt{1996ApJ...467...38K}) for elliptical galaxies.} For the relation between half-light radius and absolute magnitude in the $r^\prime$-filter, we use  equation 4 and  fitting parameters  given in table 1 from  \citet{2019RAA....19....6Z}. We utilized the $g^\prime$--$r^\prime$ derived from the SED models corresponding to each redshift to convert the observed Kron $g^\prime$-filter magnitude to the $r^\prime$-band. Using the above-mentioned relation, we compute the half-light radius for spiral  and elliptical galaxies in the  $r^\prime$-filter for an absolute magnitude in every 0.5 magnitude bin.

{Above we used the measured Kron $g^\prime$ magnitude and the template SED to calculate the required $r^\prime$ magnitude instead of the aperture $g^\prime$ and $r^\prime$, since the Kron magnitude is an appropriate measurement of the extended galaxy light. However, we will investigate below if the observed $g^\prime-r^\prime$ colour can help determine the distance of the source.} Furthermore, it is worth noting that the relation given in \citet{2019RAA....19....6Z} was developed using data from low (z$<0.2$) redshift galaxies.
As a result, the size -- luminosity relation may not hold for larger redshifts, and hence the ranges for the redshift of cNW (see below) is probably not robust. 
For completeness, we also compared the calculated magnitude with the K-correction determined using the K-calculator\footnote{See http://kcor.sai.msu.ru/} (\citealt{2010MNRAS.405.1409C}, \citealt{2012MNRAS.419.1727C}).

The resulting size--luminosity for cNW is consistent with the observed spiral galaxy size--luminosity relation for redshifts in {the range of 4.1--4.5 (Figure \ref{fig:spiral}).}
If cNW is a spiral galaxy that hosts \xa{}, the FXRT peak luminosity would thus be $\approx4\times10^{49}$ \lum. {As mentioned above the exact value of the redshift and hence luminosity is uncertain due to the large extrapolation in redshift for the observed galaxy size--luminosity relation. However, if cNW is a spiral galaxy it seems as if the source distance and hence luminosity is (too) large.}

If cNW is an elliptical galaxy host of the FXRT \xa\ instead (see Figure \ref{fig:ellip}), the observed cNW and elliptical galaxy size--luminosity correlations for the {\sc HYPERZ} template intersect for redshifts of { 0.75--0.80, 1.80--1.85, or 3.4--3.8. We found that using the {\small elliptical\textunderscore template} cNW intersects the calculated size--luminosity track for the template at two different redshift ranges namely, $z=$0.8--1.0 and 1.6--2.3. For the {\small s0\textunderscore template} cNW intersects for redshifts in the interval 1.0--1.6. Among the three elliptical (and S0) templates we used, there is quite some spread in the redshift range where the cNW size--luminosity intersects.} 

If \xa{} is related with an elliptical galaxy, the FXRT would have a peak luminosity between{ $\approx$5$ \times  10^{47}$ \lum and  3$ \times  10^{49}$ \lum.} In all these cases, a supernova SBO origin as the progenitor is excluded, and the FXRT is likely due to a beamed WD-IMBH TDE or it has a BNS merger origin.

{ A striking feature in  Figure~\ref{fig:spiral} and \ref{fig:ellip} is the loop seen at a $\log$(size) of $\sim$3.8. The location of $\log$(size)$\approx$3.8 is caused by the observed angular size of cNW combined with the angular -- physical size conversion as a function of redshift set by cosmology. For the $\Lambda$-CDM cosmology taken in this paper, the physical size for a given angular size starts to decrease as a function of redshift larger than $z\approx 1.6$. The drop in absolute magnitude from $\approx-$20.5 to $\approx-$14 for an assumed spiral galaxy (Figure~\ref{fig:spiral}) and from $\approx-$25.5 to $\approx-$18.5 for an assumed elliptical galaxy (Figure~\ref{fig:ellip}) can be explained by a combination of factors. First, our observed $g^\prime$-band magnitude is converted to an $r^\prime$-band magnitude on the basis of the $g^\prime-r^\prime$ determined from the galaxy template SED and redshift under consideration. The $g^\prime-r^\prime$ colour becomes redder for $0<z\approxlt1$ to maximise around $g^\prime-r^\prime\approx 1.5$, subsequently becoming increasingly blue from  $1\approxlt z\approxlt 2.1$, peaking at a value of $\approx -3.5$, to become redder again for even higher redshifts. Secondly, the K-correction in magnitude as a function of redshift also decreases steadily from $z\approx 1$ until a redshift of $z\approx2.2$ for our spiral galaxy template. The combined result of the $g^\prime-r^\prime$ colour and the K-correction can explain the change in the absolute magnitude of cNW for redshifts between $1\approxlt z\approxlt$2.1. Likewise, the loop seen in Figure \ref{fig:ellip} can be explained by the combined effect of these two processes, although the redshift values where the changes are largest are different due to the different spectral energy distribution of the template.}

\begin{figure}
 \flushleft
	\includegraphics[width=9cm,height=7.1cm]{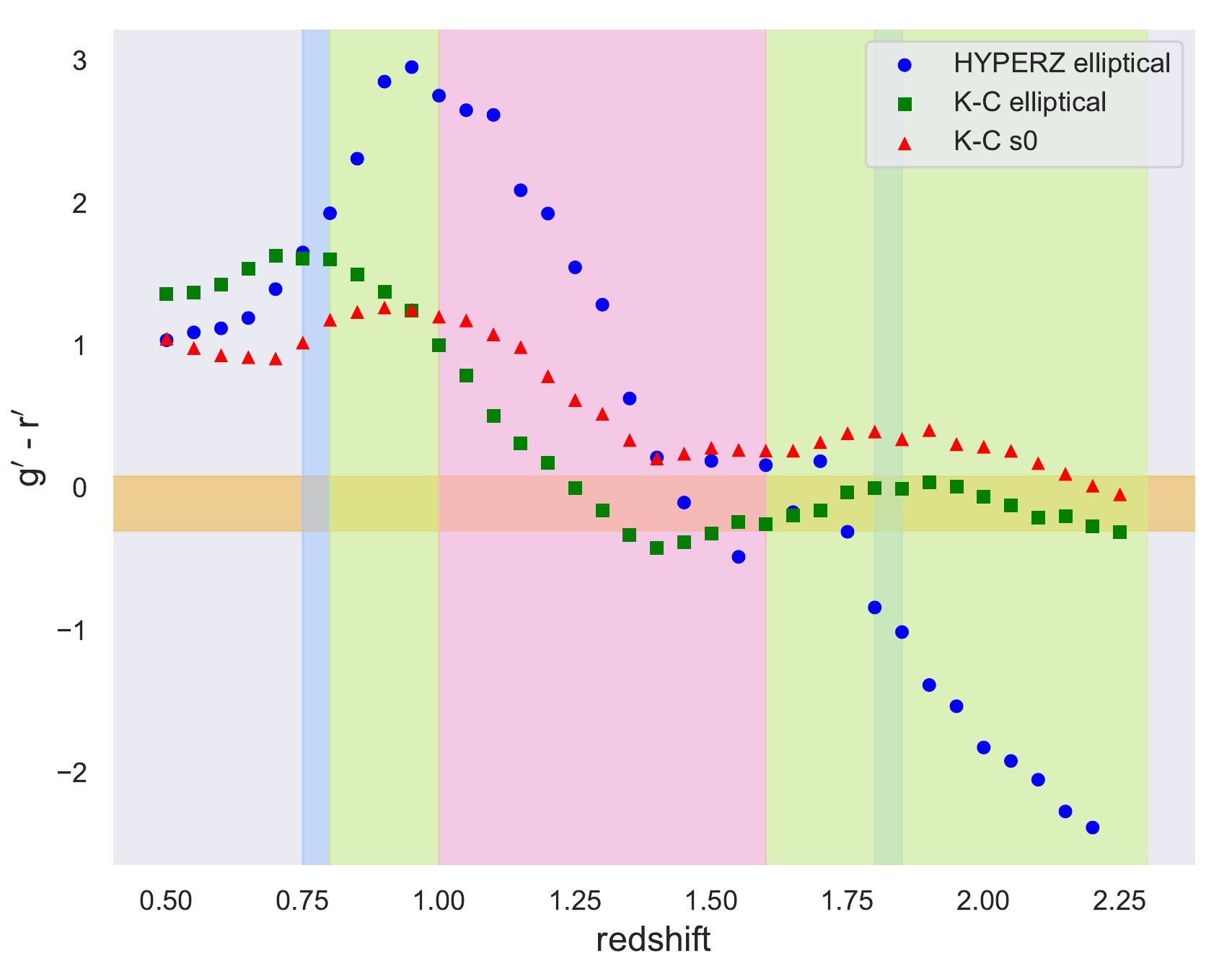}
    \caption{{The $g^\prime$--$r^\prime$ colour as a function of redshift ($z$<2.3) for the three template elliptical SEDs. Blue dots, green squares and red triangles represent the $g^\prime$--$r^\prime$ colour for the {\sc HYPERZ} {\small CWW\textunderscore E\textunderscore ext.sed}, and  the {\small elliptical\textunderscore template} and {\small s0\textunderscore template} from the {\textsc Kinney-Calzetti spectral atlas}, respectively. Vertical light blue, light green ({\small elliptical\textunderscore template}) and pink ({\small s0\textunderscore template}) shaded regions indicate the redshifts of interest for elliptical galaxies obtained using {\sc HYPERZ} and {\textsc Kinney-Calzetti spectral atlas} SEDs (see Figure \ref{fig:ellip}). Given that the redshift ranges for the {\sc HYPERZ}  and {\small elliptical\textunderscore template} overlap partially the vertical bar around redshift 1.80--1.85 has an undetermined colour. The horizontal dark yellow shaded region shows  the 1$\sigma$ colour uncertainty for the host candidate cNW, respectively. There is no redshift region where the combined colour and size--luminosity constraints overlap for the {\sc HYPERZ} and the {\small s0} {\textsc Kinney-Calzetti} templates. Only for the {\small elliptical} {\textsc Kinney-Calzetti} template the combined constraints agree to within 1$\sigma$ for a redshift in the range of 1.6--2.3}.}
    \label{g_r}
\end{figure}

{ We now come back to and use the measured aperture $g^\prime$--$r^\prime$ colour of cNW. The $g^\prime$--$r^\prime$ =$-0.11\pm0.19$ colour is inconsistent with the $g^\prime$--$r^\prime$ colours calculated from the {\sc HYPERZ} SEDs for a spiral galaxy when redshifted to $z\approx4$ at the 1$\sigma$ uncertainty level. For an elliptical galaxy, the {\sc HYPERZ}, and the two templates from the {\textsc Kinney-Calzetti spectral atlas the $g^\prime$--$r^\prime$ colour obtained for different redshifts can be compared with the measured colour of cNW and the  1$\sigma$  uncertainty on the colour as shown in Figure} \ref{g_r}. Out of the elliptical galaxy templates, only the combined constraint from the size--luminosity relations and the $g^\prime$--$r^\prime$ colour obtained from {\small elliptical\textunderscore template} and that of cNW is consistent within the 1$\sigma$ uncertainty for a redshift range of 1.6--2.3. Therefore, we deem it unlikely that cNW could be an elliptical galaxy at $z < 1.6$.} 


\subsubsection{Assuming cNW and \xa~are not associated}

Given the source density of objects as bright as or brighter than cNW around \xa\, there is an 8.3\% chance to find an object close to the FXRT at random. Therefore, we also consider the scenario where the close proximity of the source cNW and the position of the FXRT XRT 000519 arises due to chance. Here we distinguish the following two scenarios:
Assuming the FXRT is at the distance of M86, we derive an upper limit to the absolute magnitude of the host, {$M_{g_s}=-5.8$,} from the completeness limit. Given the observed absolute magnitudes of globular clusters and dwarf galaxies, part of the absolute magnitude distribution of globular clusters is not ruled out by our observations. Hence, a faint globular cluster associated with \xa{} remains a possibility (see figure 2 in \citealt{2019ARA&A..57..375S}). However, only a small fraction of the globular clusters are likely to host an IMBH, if any, and if so it is assumed that it will be the more massive globular clusters that are more likely to host an IMBH (\citealt{2020SSRv..216...39M}). This, together with the very low expected rate of white dwarf TDEs in a faint globular cluster makes that scenario unlikely.
    
If we assume that \xa~occurred at the distance of the putative galaxy group at $\sim$940 Mpc or even further away, we would derive an upper absolute magnitude limit of {$>-$14.7} from the completeness limit of  {g$^\prime=$25.2.} This absolute magnitude limit is well above the range of magnitudes seen in globular clusters (\citealt{2019ARA&A..57..375S}). However, whereas in principle we cannot rule out an association with a globular cluster in this case, the absolute magnitude constraint on the galaxy that hosts the globular cluster is stringent, still making such a scenario unlikely.


\subsection{\xb{}}
\citet{glennie} {considered it likely that }\xb{} {is associated with the galaxy cluster ACO 3581 which is at a distance of $\approx$95 Mpc} (\citealt{2005MNRAS.356..237J}). If this association is real, it implies that \xb~had a peak luminosity of $\sim$2 $\times$ $10^{44}$ \lum. Considering a distance of 95 Mpc and a completeness limit of 25.4 magnitude, we have an absolute magnitude limit of --9.5. {This implies that \xb{} cannot be associated with the brightest globular clusters. Note however that the non-detection of a galaxy to which such a globular cluster would belong makes this scenario unlikely. Also, if globular clusters host IMBHs it will be those at this bright massive end of the globular clusters that do (\citealt{2018ApJ...856...92F}). Hence, \xb{} is unlikely to be linked to a WD-IMBH TDE in a globular cluster.}
\vspace{-0.1 cm}

\section{Conclusions}
We present a search in the optical for the host of two fast X-ray transients. For one, \xa, we detected an extended source in the GTC/HiPERCAM $g_s$- and $r_s$-band images with a $g_s$-band Kron magnitude of {${\sim}$~26.29$\pm$0.09}. We consider this source, so called cNW, a candidate host galaxy for \xa. {Under the assumption that \xa{} originated in cNW, if at the distance of M86, we cannot completely rule out cNW being a dwarf galaxy. If cNW lies at the distance of the putative galaxy group at a distance of 940.2 $\pm$ 0.4 Mpc that we identified, the fast X-ray transient could originate from a blue dwarf galaxy.} We also used  galaxy size--luminosity relations for spiral or elliptical galaxies to constrain the redshift of cNW. {If cNW lies further away, 
it could be an elliptical galaxy at a redshift in the range of 1.6--2.3, comparing the colour of cNW with that of redshifted template galaxy colours.} For all these host scenarios it is effectively ruled out that \xa\ is due to a supernova shock break-out signal, in line with the non-detection of a supernova.

Even if the close proximity of the host candidate and the position of the FXRT \xa~is due to a chance alignment, { we cannot definitively rule out a dwarf galaxy at the distance of M86, although it is disfavoured.}

The non-detection of a host candidate for \xb{} does not allow us to put stringent constraints on the origin of this event. { However, we consider it unlikely that \xb{} is associated with a source in a globular cluster.}

 \section*{Acknowledgements}
Based (in part) on observations collected at the European Southern Observatory under ESO programme 0100.B-0323. The Pan-STARRS1 Surveys (PS1) and the PS1 public science archive have been made possible through contributions by the Institute for Astronomy, the University of Hawaii, the Pan-STARRS Project Office, the Max-Planck Society and its participating institutes, the Max Planck Institute for Astronomy, Heidelberg and the Max Planck Institute for Extraterrestrial Physics, Garching, The Johns Hopkins University, Durham University, the University of Edinburgh, the Queen's University Belfast, the Harvard-Smithsonian Center for Astrophysics, the Las Cumbres Observatory Global Telescope Network Incorporated, the National Central University of Taiwan, the Space Telescope Science Institute, the National Aeronautics and Space Administration under Grant No. NNX08AR22G issued through the Planetary Science Division of the NASA Science Mission Directorate, the National Science Foundation Grant No. AST-1238877, the University of Maryland, Eotvos Lorand University (ELTE), the Los Alamos National Laboratory, and the Gordon and Betty Moore Foundation.Funding for the Sloan Digital Sky 
Survey IV has been provided by the 
Alfred P. Sloan Foundation, the U.S. 
Department of Energy Office of 
Science, and the Participating 
Institutions. 

SDSS-IV acknowledges support and 
resources from the Center for High 
Performance Computing  at the 
University of Utah. The SDSS 
website is www.sdss.org. SDSS-IV is managed by the 
Astrophysical Research Consortium 
for the Participating Institutions 
of the SDSS Collaboration including 
the Brazilian Participation Group, 
the Carnegie Institution for Science, 
Carnegie Mellon University, Center for 
Astrophysics | Harvard \& 
Smithsonian, the Chilean Participation 
Group, the French Participation Group, 
Instituto de Astrof\'isica de 
Canarias, The Johns Hopkins 
University, Kavli Institute for the 
Physics and Mathematics of the 
Universe (IPMU) / University of 
Tokyo, the Korean Participation Group, 
Lawrence Berkeley National Laboratory, 
Leibniz Institut f\"ur Astrophysik 
Potsdam (AIP),  Max-Planck-Institut 
f\"ur Astronomie (MPIA Heidelberg), 
Max-Planck-Institut f\"ur 
Astrophysik (MPA Garching), 
Max-Planck-Institut f\"ur 
Extraterrestrische Physik (MPE), 
National Astronomical Observatories of 
China, New Mexico State University, 
New York University, University of 
Notre Dame, Observat\'ario 
Nacional / MCTI, The Ohio State 
University, Pennsylvania State 
University, Shanghai 
Astronomical Observatory, United 
Kingdom Participation Group, 
Universidad Nacional Aut\'onoma 
de M\'exico, University of Arizona, 
University of Colorado Boulder, 
University of Oxford, University of 
Portsmouth, University of Utah, 
University of Virginia, University 
of Washington, University of 
Wisconsin, Vanderbilt University, 
and Yale University.

The design and construction of HiPERCAM was funded by the European Research Council under the European Union’s Seventh Framework Programme (FP/2007-2013) under ERC-2013-ADG Grant Agreement no. 340040 (HiPERCAM). VSD and HiPERCAM operations are supported by STFC grant ST/V000853/1. This work is based on observations made with the Gran Telescopio Canarias (GTC), installed at the Spanish Observatorio del Roque de los Muchachos of the Instituto de Astrofísica de Canarias, in the island of La Palma. We would like to thank the staff of the GTC for their continued support for HiPERCAM in what have been very difficult circumstances over the past year.

We thank Tom Marsh for the use of {\sc molly}. DMS acknowledges support from the ERC under the European Union’s Horizon 2020 research and innovation programme (grant agreement no. 715051; Spiders). DMS acknowledges the Fondo Europeo de Desarrollo Regional (FEDER) and the Canary Islands government for the financial support received in the form of a grant with number PROID2020010104. F.O. acknowledges the support of the GRAWITA/PRIN-MIUR project: "\textit{The new frontier of the Multi-Messenger Astrophysics: follow-up of electromagnetic transient counterparts of gravitational wave sources}". ZKR acknowledges funding from the Netherlands Research School for Astronomy (NOVA). KM is funded by EU H2020 ERC grant no. 75863. MF acknowledges the support of a Royal Society - Science Foundation Ireland University Research Fellowship. ANID grants Programa de Capital Humano Avanzado folio 21180886 (JQ–V), CATA-Basal AFB-170002 (JQ–V), FONDECYT Regular 1190818, 1200495 (JQ–V) and Millennium Science Initiative ICN12 009 (JQ–V). MAPT acknowledges support from the State Research Agency (AEI) of the Spanish Ministry of Science, Innovation and Universities (MCIU) and the European Regional Development Fund (FEDER) under grant AYA2017-83216-P and via a Ram\'on  y Cajal Fellowship (RYC-2015-17854). This work is part of the research programme Athena with project
number 184.034.002, which is financed by the Dutch Research
Council (AI). SJB would like to thank their support from Science Foundation Ireland and the Royal Society (RS-EA/3471). D.E acknowledges discussions with Daniele Malesani, Davide Lena, Patrick Rauer and Kristhell Lopez. We thank the anonymous referee for the helpful comments on this manuscript.

\section*{DATA AVAILABILITY}
All data will be made available in a reproduction package uploaded
to Zenodo.

\small
\bibliographystyle{mnras}
\bibliography{mybibfile}

\end{document}